\documentclass[twocolumn,showpacs,prb,amsmath,amssymb]{revtex4}
\usepackage{amsfonts}

\usepackage{graphicx}
\usepackage{dcolumn}
\usepackage{bm}
\usepackage{epsfig}
\usepackage{amsmath}
\usepackage[dvips]{color}
\usepackage{booktabs,threeparttable}

\newcommand{\beann}{\begin{eqnarray}} \newcommand{\eeann}{\end{eqnarray}}
\newcommand{\bea}{\begin{eqnarray}} \newcommand{\eea}{\end{eqnarray}}

\begin{document}

\title{DFT-based calculation of Coulomb blockade in molecular junction}
\author{Bo~Song}
\email{bo.song@tu-dresden.de}
\affiliation{Institute~for~Materials~Science,~Dresden~University~of~Technology,~D-01062~Dresden,~Germany}
\date{\today}

\begin{abstract}
Quantum transport through single molecules is very sensitive to the
strength of the molecule-electrode contact. When a molecular junction
weakly coupled to external electrodes, charging effects do play an
important role (Coulomb blockade regime). In this regime, the non-equilibrium
Green function is usually substituted with master equation approaches,
which prevents the density functional theory from describing Coulomb blockade in
non-equilibrium case. Last year, we proposed an Ansatz to combine the non-equilibrium
Green function technique with the equation of motion method. With help of it,
Coulomb blockade was obtained by non-equilibrium Green function,
and completely agrees with the master equation results [Phys.~Rev.~B~\textbf{76},~045408~(2007)].
Here, by the Ansatz, we show a new way to introduce Coulomb blockade correction to DFT calculation
in non-equilibrium case. And the characteristics of Coulomb blockade are obtained
in the calculation of a $toy$ molecule correctly.
\end{abstract}

\maketitle

\section{introduction}

%
Single molecule electronics\cite{NR03,JGA00,cfr05} has been
mostly investigated in the high temperature and strong contact to
the electrode regime. The opposite limit of low temperature and
weakly coupled molecular junctions poses a challenge to the
currently available experimental techniques. Still the possibility
to probe the spectroscopy of single molecule junctions via a lateral
gate could offer new insights to the peculiar coupling of the
electrical and mechanical degrees of freedom at the nanoscale.
%
In order to be able to establish the transport mechanisms
governing such molecular junctions,
a technique which could tackle on one hand single electron
charging effects and, on the other hand, the inclusion of the
electron-vibron coupling is of extreme importance.

In the last ten years, the nonequilibrium Green function (NEGF)
formalism has been successfully employed to describe transport
observables on the base of a density functional theory (DFT)
description of the electronic
structure\cite{cfr05,transiesta02,TaylorBS03,gdftb02,
gdftb02-2,gdftb04-rpp,gdftb04-nl,smeago06,abinitioNEGF05,KBY03} and model Hamiltonian
approaches,\cite{GalperinN03,GalperinNR06,Pals96jpcm}
%
Recently it is applied for the influence
of the vibron dynamics onto a molecular transistor, and lots of
excellent results are obtained.\cite{Ryndyk05prb,Ryndyk06prb,e-ph_gdftb}


%
However, when coming to the CB regime, the NEGF
method has been usually substituted with master equation
approaches (ME), which prevents DFT from describing CB effects.
Last year, we proposed an Ansatz to combine NEGF
with equation of motion (EOM) method.\cite{BoSONG_M-CB_06-07} With help of the Ansatz above,
in non-equilibrium case, the Coulomb
blockade can be completely described just within single-particle space simply,
and the result fully agrees with the one from ME which is performed in many-particle space.
%

For DFT to describe Coulomb blockade,\cite{KBY03,Palacios05} the
double subspace (spin-up and spin-down) is usually employed.
However, it is hard to obtain the Coulomb
blockade effects correctly in non-equilibrium case.
Our purpose is to introduce Coulomb blockade effect to non-equilibrium
DFT calculation by this Ansatz.

In this paper, with the help of the Ansatz in
Ref.~\cite{BoSONG_M-CB_06-07}, by the model Hamiltonian and EOM
approach, we propose a self energy to describe CB effects in
non-equilibrium case. Electronic occupation number,
electronic current and differential conductance are calculated.
The charging-induced steps and the ratio of 2/3~:~1/3 in the step heights
of the occupation number and the current,
which are the important characteristics
of CB by ME,\cite{Datta_ratio2/3-1/3} is obtained correctly. The
comparison with the complete CB results\cite{BoSONG_M-CB_06-07} also
is done. For the occupation number and the current, the difference
is very small. For the conductance, only in the peak height, the
difference is very clear, while there is no difference in the peak
position. The CB stability diagram is also shown in the paper. We can see
that the self energy can describe CB characteristics.
The more important is that it is very convenient to introduce this
self energy to DFT code even in non-equilibrium case. A scheme to
perform CB correction in DFT calculation is suggested with double counting
correction (DCC). Then it is realized in gDFTB.\cite{gdftb02,gdftb04-rpp,gdftb04-nl} A
\textit{toy} molecule is taken for testing, and the CB
characteristics are shown in the results correctly.

\begin{figure}[b]\label{toy_mol}
\begin{center}
\includegraphics[width=3.0cm,angle=-90]{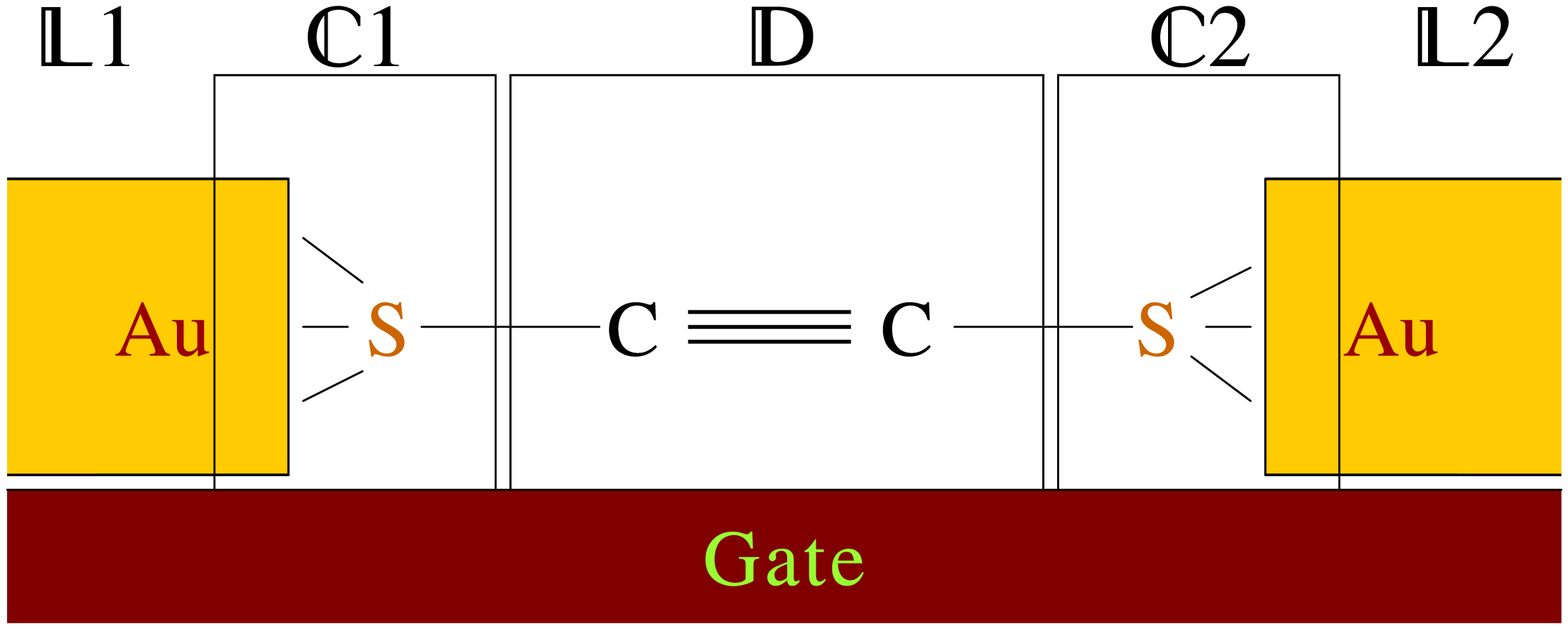}
\caption{(Color online) The molecule for Coulomb blockade
calculation. See text for details.}
\end{center}
\end{figure}

The paper is organized as follows: firstly a self energy for CB is
proposed by the model Hamiltonian and EOM approach (Sec.~II);
secondly, based on the self energy above, a scheme with DCC is
proposed to introduce CB correction to DFT code (Sec.~IIIA); finally, the
calculation on a \textit{toy} molecule is performed in weak coupling
regime, and the CB characteristics are shown in the results correctly (Sec.~IIIB).

\section{method and formula}

\begin{figure}[t]\label{N-I-V-comp}
\begin{center}
\includegraphics[width=1.0\hsize]{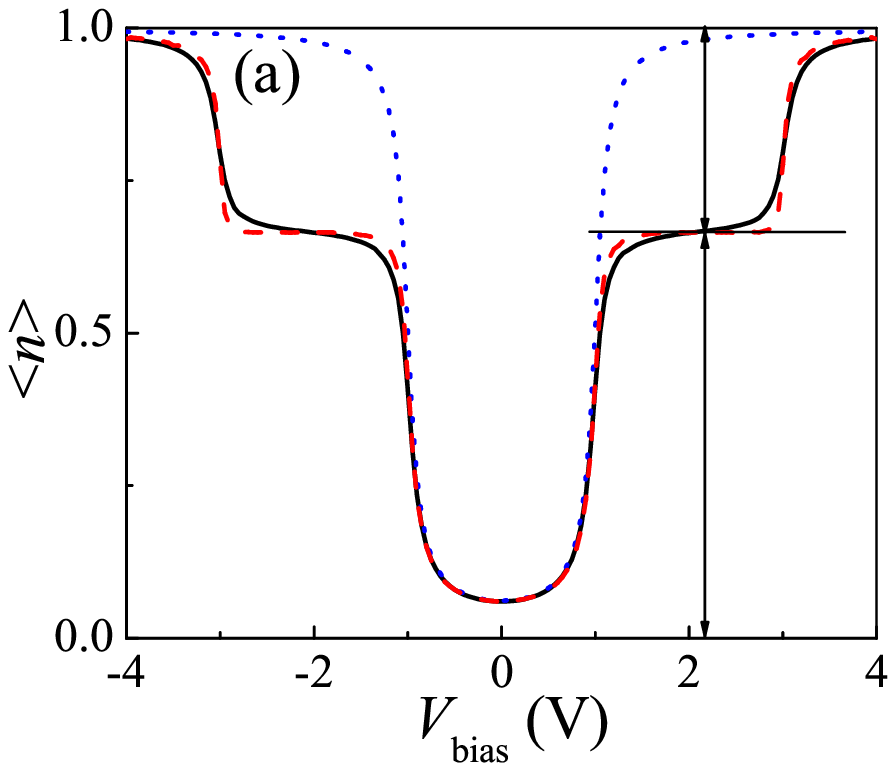}
\includegraphics[width=1.0\hsize]{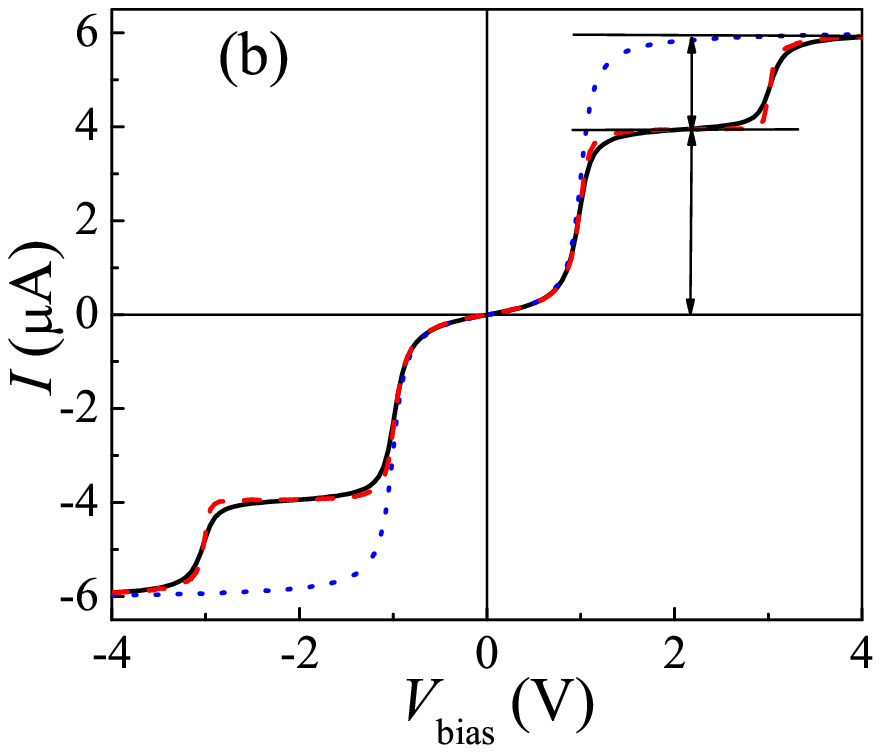}
\caption{(Color online) (a) the comparison on the occupation number
of electrons as a function of bias voltage. (b) the comparison on
the current as a function of bias voltage.
$\epsilon^{0}_{\uparrow}=\epsilon^{0}_{\downarrow}=-0.5~e$V,
$V_{\textrm{g}}=1.0~$V, $U=1.0~e$V, $\Gamma=0.05~e$V,
$V_{\textrm{L}}=-V_{\textrm{R}}=V_{\textrm{bias}}/2$. $\langle
n\rangle=\langle n_{\uparrow}\rangle+\langle n_{\downarrow}\rangle$.
The red dash curves are from approximation in Eq.~\ref{GF_r/a_2},
the black solid curves from
the truncation in Ref.~\cite{BoSONG_M-CB_06-07} which fully agree with the results from ME,~\cite{BoSONG_M-CB_06-07}
while the blue dot curves are for the case
$U=0~e$V in which there is no CB. In the red dash curves and the black solid curves,
the new steps appear for charging effect,
and the heights of them are in the ratio of 2/3~:~1/3,
which are the important characteristics of CB obtained by ME.~\cite{Datta_ratio2/3-1/3}}
\end{center}
\end{figure}

In non-equilibrium DFT calculation here, the CB correction will be introduced to each level
of the molecular fragment $\mathbb{D}$ (see Fig.~1).~\cite{note_a}
Therefore, the multi-level Anderson impurity model is read as follows,
\begin{eqnarray}\label{H}
H&=&H_{\textrm{D}}+\sum_{\alpha}(H_{\alpha}+H_{\alpha \textrm{D}}),
\end{eqnarray}
with
\begin{eqnarray}\label{H-D}
&&H_{\textrm{D}}=\sum_{m,\sigma}{\epsilon^{0}_{m,\sigma}d^{\dagger}_{m,\sigma}d^{\phantom{\dagger}}_{m,\sigma}}
+\frac{1}{2}\sum_{m,\sigma}{U_{m}n_{m,\sigma}n_{m,\bar{\sigma}}},\\
&&H_{\alpha}=\sum_{k,\sigma}\epsilon^{\alpha}_{k,\sigma}c^{\dagger}_{k,\sigma,\alpha}
c^{\phantom{\dagger}}_{k,\sigma,\alpha},\\
&&H_{\alpha,\textrm{D}}=\sum_{k,m,\sigma}(V^{\phantom{\ast}}_{\alpha,k,m,\sigma}c^{\dagger}_{k,\sigma,\alpha}
d^{\phantom{\dagger}}_{m,\sigma}+h.c.),
\end{eqnarray}
where $d$ and $c$ are the operators for electrons on the dot and on the left ($\alpha=\textrm{L}$) and the right
($\alpha=\textrm{R}$) lead, $U_{m}$ is the charging energy of level $m$, $\epsilon_{m,\sigma}$
is the ($m$, $\sigma$) level of the quantum dot, while $\epsilon^{\alpha}_{k,\sigma}$ is the
spin $\sigma$ level of lead $\alpha$ in $k$ space, $\sigma=\uparrow, \downarrow$.
With the help of the EOM and the truncation approximation, we can obtain a closed set of equations
for the retarded and advanced GFs $G^{r/a}_{m,\sigma;n,\tau}$,
\begin{eqnarray}\label{GF_r/a_1}
&&(\omega-\epsilon^{0}_{m,\sigma}-\Sigma^{(\alpha)r/a}_{m,\sigma})G^{r/a}_{m,\sigma;n,\tau}
=\delta_{m,n}\delta_{\sigma,\tau}\nonumber\\&&\hspace{4cm}+U_{m}G^{(2)r/a}_{m,\sigma;n,\tau},\\
\label{GF_r/a_2}
&&(\omega-\epsilon^{0}_{m,\sigma}-U_{m}\pm\textrm{i}\eta)G^{(2)r/a}_{m,\sigma;n,\tau}
=\langle n_{m,\bar{\sigma}}\rangle\delta_{m,n}\delta_{\sigma,\tau}\nonumber\\
&&\hspace{2.365cm}+\Sigma^{(\alpha)r/a}_{m,\sigma}\langle
n_{m,\bar{\sigma}}\rangle G^{r/a}_{m,\sigma;n,\tau},
\end{eqnarray}
where
\begin{eqnarray}
&&G^{r/a}_{m,\sigma;n,\tau}=\langle\langle d^{\phantom{\dagger}}_{m,\sigma}|d^{\dagger}_{n,\tau}\rangle\rangle^{r/a},\\
&&G^{(2)r/a}_{m,\sigma;n,\tau}=\langle\langle n^{\phantom{\dagger}}_{m,\bar{\sigma}}
d^{\phantom{\dagger}}_{m,\sigma}|d^{\dagger}_{n,\tau}\rangle\rangle^{r/a},
\end{eqnarray}
and
\begin{eqnarray}\label{self-E}
\Sigma^{(\alpha)r/a}_{m,\sigma}(\omega)=\Sigma^{\textrm{L},r/a}_{m,\sigma}+\Sigma^{\textrm{R},r/a}_{m,\sigma}
=\sum_{\alpha,k}\frac{|V_{\alpha,k,m,\sigma}|^{2}}{\omega-\epsilon^{\alpha}_{k,\sigma}\pm \textrm{i}\eta}
\end{eqnarray}
are the electron self-energies from leads with $\eta=0^{+}$.

Re-arranging Eqs.~(\ref{GF_r/a_1}) and (\ref{GF_r/a_2}), we can obtain the retarded GF as follows,
\begin{eqnarray}\label{GF-r}
G^{r}=G^{(U)r}+G^{(U)r}\Sigma^{(\alpha)r}G^{r},
\end{eqnarray}
with
\begin{eqnarray}\label{GF-u-r}
&&G^{(U)r}=G^{(0)r}+G^{(0)r}\Sigma^{r}_{\textrm{H}}G^{(1)r},\\
&&G^{(0)r}=\left\{\omega-H_{0}+\textrm{i}\eta\right\}^{-1},\\
&&G^{(1)r}=\{\omega-H_{0}+U+\textrm{i}\eta\}^{-1},
\end{eqnarray}
$H_{0}$ is a diagonal matrix composed of $\epsilon^{0}_{m,\sigma}$,
$\Sigma^{r}_{\textrm{H}}(=\Sigma^{a}_{\textrm{H}})$ is the one of $\langle n_{m,\bar{\sigma}}\rangle U_{m}$,
and $U$ is of $U_{m}$,
while $\Sigma^{(\alpha)r}$ is the self-energy matrix from Eq. (\ref{self-E}).

From Eq. (\ref{GF-r}), we can see that the Coulomb interaction is just included in $G^{(U)r}$.
Therefore, with the help of Eq. (\ref{GF-u-r}), the retarded CB self energy $\Sigma^{(\textrm{CB})r}$
can be obtained by the relation
\begin{eqnarray}\label{eq:Key-0}
[G^{(U)r}(\omega)]^{-1}&=&\left\{\frac{1}{\omega-H_{0}+\textrm{i}\eta}
+\frac{1}{\omega-H_{0}+\textrm{i}\eta}\right.\nonumber\\
&&\left.\cdot\Sigma^{r}_{\textrm{H}}\cdot\frac{1}{\omega-H_{0}-U+\textrm{i}\eta}\right\}^{-1}\nonumber\\
&\equiv&\left\{\omega-H_{0}-\Sigma^{(\textrm{CB})r}\right\}^{-1}.
\end{eqnarray}
The result is as follows,
\begin{eqnarray}\label{eq:Key-1}
\Sigma^{(\textrm{CB})r}=\omega-H_{0}-[G^{(U)r}]^{-1}.
\end{eqnarray}

\begin{figure}\label{G-V-comp}
\begin{center}
\epsfxsize=1.0\hsize \epsfbox{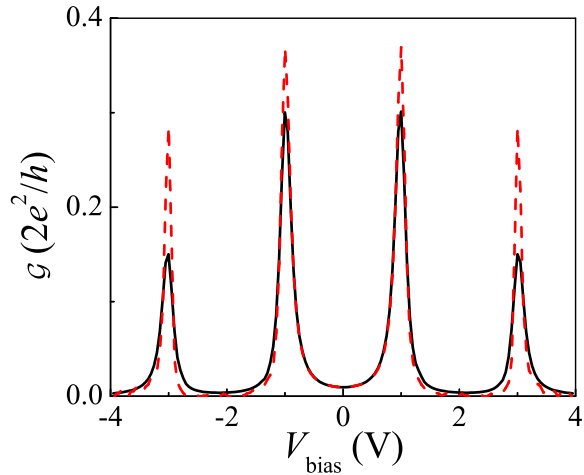} \caption{(Color online)
The comparison on conductance with the same parameters in Fig.~1.
The red dash curve is from approximation in Eq.~\ref{GF_r/a_2},
while the black solid curve from
the truncation in Ref.~\cite{BoSONG_M-CB_06-07} which fully agrees with the one from
ME.~\cite{BoSONG_M-CB_06-07}}
\end{center}
\end{figure}

By the Ansatz in Ref.~\cite{BoSONG_M-CB_06-07}, the lesser GF can be
written out directly,
\begin{eqnarray}
G^{<}&=&G^{(U)<}+G^{(U)<}\Sigma^{(\alpha)a}G^{a}\nonumber\\
&+&G^{(U)r}\Sigma^{(\alpha)<}G^{a}+G^{(U)r}\Sigma^{(\alpha)r}G^{<},
\end{eqnarray}
with
$\Sigma^{(\alpha)<}=\textrm{i}\sum_{\alpha}{\Gamma_{\alpha}f_{\alpha}(\omega)}$,
and
$\Gamma_{\alpha}=\textrm{i}(\Sigma^{r}_{\alpha}-\Sigma^{a}_{\alpha})$,
$f_{\alpha}(\omega)=f(\omega-\mu_{\alpha})$, while
$\Sigma^{r/a}_{\alpha}$ are the diaganol matrix composed of
$\Sigma^{\alpha,r/a}_{m,\sigma}$, and $f$ is the equilibrium Fermi
function. After the re-arrangement by the way in Ref.~\cite{HJ96}, we can obtain,
\begin{eqnarray}\label{GF-l}
G^{<}&=&G^{r}\Sigma^{<}G^{a},\\
\label{Sigma-l}
\Sigma^{<}&=&\Sigma^{(\alpha)<}+\Sigma^{(\textrm{CB})<}.
\end{eqnarray}
By the helps of $\Sigma^{(\textrm{CB})<}\rightarrow 0$ (see appendix
\ref{app:Sigma-CB-l}), we could get $\Sigma^{<}=\Sigma^{(\alpha)<}$
if $\Sigma^{(\alpha)<}\ne 0$.

Therefore, here, the current $I$ can be calculated simply by the
Landaur formula,~\cite{Landaur89,datta_book95}
\begin{eqnarray}\label{Landaur-formula}
I=\frac{2e}{h}\int{d\omega}~\textrm{Tr}
\left\{\Gamma_{\textrm{L}}G^{r}\Gamma_{\textrm{R}}G^{a}\right\}
\cdot\left[f_{\textrm{L}}(\omega)-f_{\textrm{R}}(\omega)\right],
\end{eqnarray}
instead of the complicated formula in
Ref.~\cite{mw92,BoSONG_M-CB_06-07}.

\begin{figure}\label{G-V-V}
\begin{center}
\includegraphics[width=8.0cm]{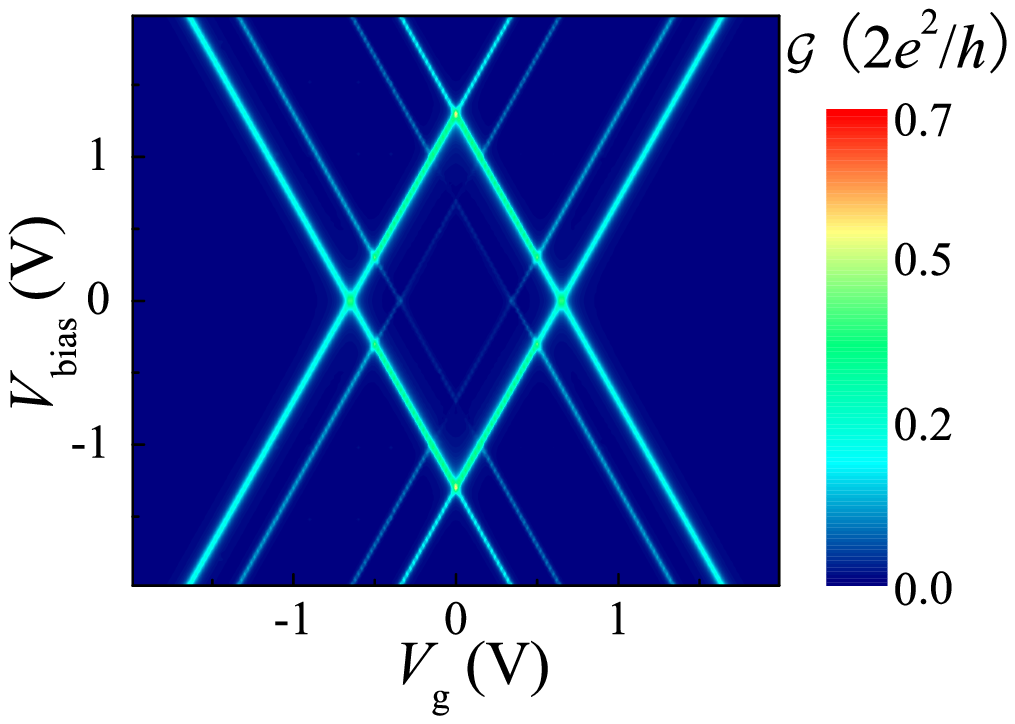}
\caption{(Color online) The
CB stability diagram (the contour plot of the differential conductance)
calculated by approximation in Eq.~\ref{GF_r/a_2}, with
$\epsilon^{0}_{\uparrow}=-0.65~e$V,
$\epsilon^{0}_{\downarrow}=-0.45~e$V, $U=1.0~e$V, $\Gamma=0.02~e$V.}
\end{center}
\end{figure}

In the case of double levels ($m\equiv 1$ and
$\epsilon^{0}_{1,\uparrow}=\epsilon^{0}_{1,\downarrow}$), the
numerical calculation is performed, which contributes the direct
comparison between the truncation in Eq.~(\ref{GF_r/a_2}) and the
one in Ref.~\cite{BoSONG_M-CB_06-07}
by which CB results fully in agreement with the ones from ME can be obtained.~\cite{BoSONG_M-CB_06-07}
The comparisons of the electronic occupation number
$\langle n\rangle$ and the current $I$ as
a function of the bias voltage $V_{\textrm{bias}}$ at fixed gate
voltage $V_{\textrm{g}}$ are shown in Fig.~2.
Firstly, the difference of the results by the two methods is very
small. Secondly, the CB characteristics are very clear in them:
the steps appear for charging-induced level-split,
and the step heights are in the ratio of 2/3~:~1/3,
which are also obtained by ME as the important
CB characteristics.~\cite{Datta_ratio2/3-1/3}
Only in the comparison of differential conductance ($\mathcal{G}=\partial
I/\partial V_{\textrm{bias}}$) (see Fig.~3), the difference in
the heights of the peaks ($\mathcal{G}$-axis) is very clear,
while no difference appears in the positions of the peaks
($V_{\textrm{bias}}$-axis). From the CB stability diagram
in the case of $m\equiv 1$ and
$\epsilon^{0}_{1,\uparrow}\ne\epsilon^{0}_{1,\downarrow}$ (Fig.~4),
it can be seen that in this case, the approximation in Eq.~(\ref{GF_r/a_2})
\textit{almost} includes all CB characteristics (the
\textit{complete} CB stability diagram by NEGF
is shown in
Ref.~\cite{BoSONG_M-CB_06-07}).

It should be noted that although the approximation above can just include \textit{some}
CB characteristics correctly,~\cite{note_a} it is very convenient to be introduced to DFT calculation
(which will be shown in the next chapter).

\section{scheme for DFT}

\subsection{scheme}
For CB calculation, the system is partitioned as follows
(shown in Fig.~1): the molecular fragment $\mathbb{D}$ is the CB
part, the fragments $\mathbb{C}1$ and $\mathbb{C}2$ are contacting
area, and the $\mathbb{L}1$ and $\mathbb{L}2$ are the leads. Then
we just introduce CB correction in fragment $\mathbb{D}$,
while the non-equilibrium calculation should
be performed within the fragment
$\mathbb{C}1-\mathbb{D}-\mathbb{C}2$.

In DFT, the KS equation $H^{\textrm{KS}}\Psi=SE\Psi$
can be re-written as,
\begin{eqnarray}
S^{-1}H^{\textrm{KS}}\Psi=E\Psi,
\end{eqnarray}
with $S$ is the overlap matrix.
Then an effective Hamiltonian matrix\cite{Newton91CR,Ratner05JACS}
for model-Hamiltonian calculation can be obtained from KS one,
\begin{eqnarray}
H^{\textrm{eff}}=S^{-1}H^{\textrm{KS}}.
\end{eqnarray}

After performing the transformation on
$H^{\textrm{eff}}_{\mathbb{C}1-\mathbb{D}-\mathbb{C}2}$ from atomic
basis to the fragment basis (which are from the eignvectors of
molecular fragment $\mathbb{C}1-\mathbb{D}-\mathbb{C}2$ and are
orthonormal), we take,
\begin{equation}
h^{\textrm{CB}}_{\mathbb{C}1-\mathbb{D}-\mathbb{C}2,m,n}=
\left\{
\begin{aligned}
&h^{\textrm{eff}}_{\mathbb{C}1-\mathbb{D}-\mathbb{C}2,m,m},&(m=n),\\
&0,&(m\ne n),
\end{aligned}
\right.
\end{equation}
with $m,n$ are the index for the eigenvectors of the fragment
$\mathbb{C}1-\mathbb{D}-\mathbb{C}2$,
and $h^{\textrm{eff}}_{\mathbb{C}1-\mathbb{D}-\mathbb{C}2}$
are the the element of effective Hamiltonian matrix $H^{\textrm{eff}}$.

Within the fragment basis, from equation~(\ref{eq:Key-1}), we can obtain
the self energy for CB correction in DFT as follows,
\begin{equation}
\widetilde{\Sigma}^{(\textrm{CB})r}_{m,m,\sigma}=(\widetilde{\omega}-\epsilon^{0}_{m,\sigma})
-\frac{(\widetilde{\omega}-\epsilon^{0}_{m,\sigma})*(\widetilde{\omega}-\epsilon^{0}_{m,\sigma}-U_{m})}
{\widetilde{\omega}-\epsilon^{0}_{m,\sigma}-(1-\langle
n_{m,\bar{\sigma}}\rangle)\cdot U_{m}},
\end{equation}
where $\widetilde{\omega}=\omega+\textrm{i}\eta$, $U_{m}$ is the
Hubbard energy of the fragment orbital \textit{m}, and
$\widetilde{\Sigma}^{(\textrm{CB})r}_{m,n,\sigma}=0$ if $m\ne n$. The occupation
number of electrons $\langle n_{m,\bar{\sigma}}\rangle=\rho_{m,m,\bar{\sigma}}$
can be obtained by the transformation on the density matrix from
atomic basis $\rho_{\mu,\nu,\bar{\sigma}}$ to fragment basis
$\rho_{m,n,\bar{\sigma}}$. Considering DCC along the idea similar to
the case in LDA+U,~\cite{dcc97,dcc05} we can get
$\epsilon^{0}_{m,\sigma}=h^{\textrm{CB}}_{\mathbb{C}1-\mathbb{D}
-\mathbb{C}2,m,m,\sigma}-\langle
n_{m,\bar{\sigma}}\rangle~U_{m}$. The way to calculate $U_{m}$
within DFTB/gDFTB is shown in appendix \ref{app:U}.

Finally the CB-correction self energy $\widetilde{\Sigma}^{(\textrm{CB})r}$
will be transformed back to the atomic basis
$\widetilde{\Sigma}^{(\textrm{CB})r}_{\mu,\nu,\sigma}$ from the fragment basis
$\widetilde{\Sigma}^{(\textrm{CB})r}_{m,m,\sigma}$. By the help of
Eqs.~(\ref{GF-r}),~(\ref{eq:Key-0}),~(\ref{GF-l})~and~(\ref{Sigma-l}),
introducing the self energy $\Sigma^{(\alpha)r/a/<}$ from leads, we
can calculate the GFs as follows,
\begin{equation}
G^{r/a}_{\mathbb{C}1-\mathbb{D}-\mathbb{C}2}=\left\{\omega S-H^{\textrm{KS}}_{\mathbb{C}1-\mathbb{D}-\mathbb{C}2}
-\Sigma^{(\alpha)r/a}-\widetilde{\Sigma}^{(\textrm{CB})r/a}
\right\}^{-1},
\end{equation}
and
\begin{equation}
G^{<}_{\mathbb{C}1-\mathbb{D}-\mathbb{C}2}=G^{r}_{\mathbb{C}1-\mathbb{D}-\mathbb{C}2}\Sigma^{(\alpha)<}
G^{a}_{\mathbb{C}1-\mathbb{D}-\mathbb{C}2}.
\end{equation}

It should be noted that the eigenvectors of the fragment
$\mathbb{C}1-\mathbb{D}-\mathbb{C}2$ is updated in every cycle
according to the updated KS Hamiltonian.

\begin{figure}\label{fig:tunneling}
\begin{center}
\epsfxsize=0.93\hsize \epsfbox{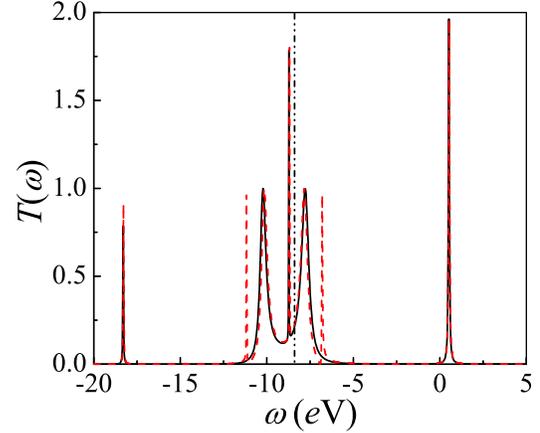} \caption{(Color
online) The transmission function $T(\omega)$ in spectral space
$\omega$. The black solid curve is for the transmission function in the
case without CB correction, and the red dash is with CB correction and $U_{m}\equiv
1.0~e\textrm{V}$, while the black dot-dash line is the Fermi level
$E_{\textrm{F}}=-8.5~e\textrm{V}$. In the case without CB correction, we
can see that there are seven levels close to the Fermi energy, the
HOMO and the LUMO+1 are double-degenerate, respectively. When
CB correction is introduced, for charging effect,
the LUMO and the HOMO-2 will split, respectively. See text for details. }
\end{center}
\end{figure}

\subsection{calculation on a \textit{toy} molecule}

\begin{figure}\label{fig:occ-cur-gDFTB}
\begin{center}
\includegraphics[width=1.0\hsize]{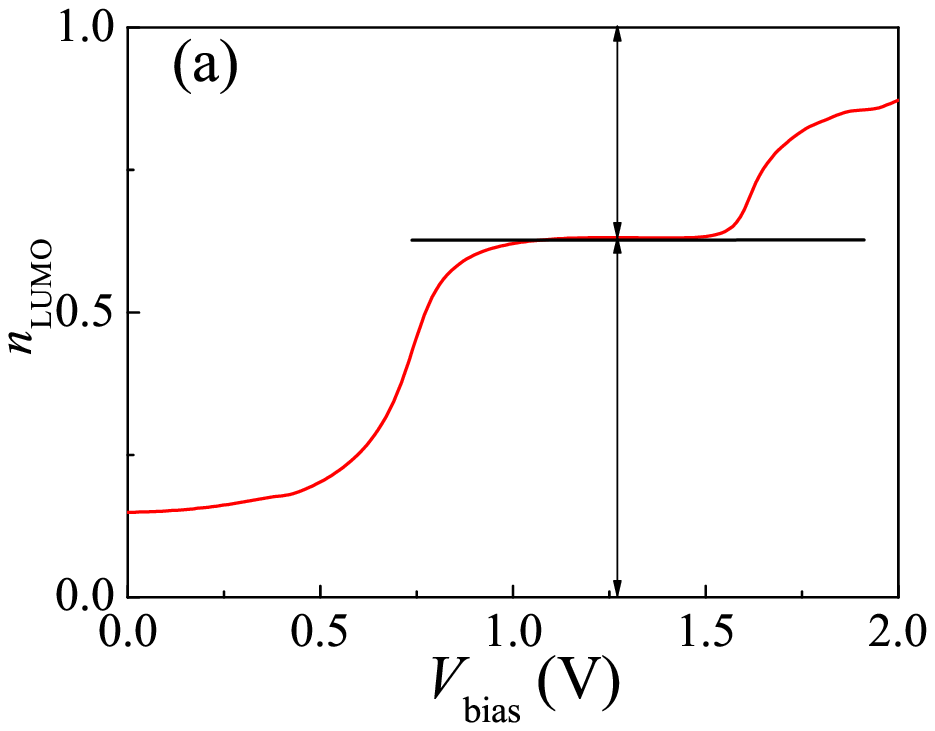}
\includegraphics[width=1.0\hsize]{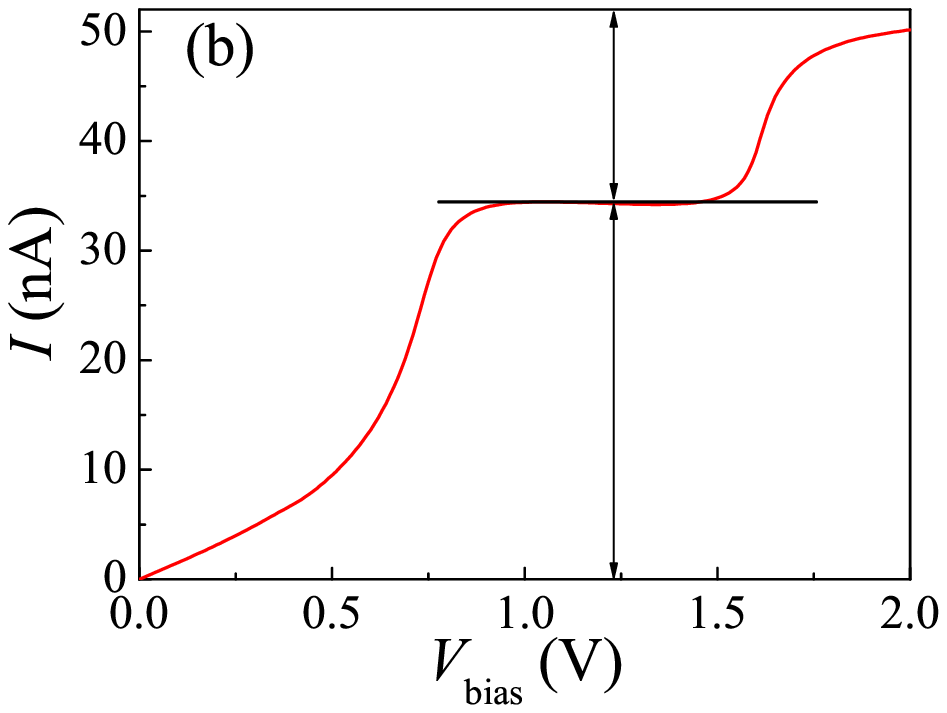}
\caption{(Color online) (a) the electronic occupation number $n_{\textrm{LUMO}}$ of LUMO
 as a function of the bias voltage. (b) the current
$I$ as a function of the bias voltage. In them, the CB characteristics
(the charging-induced steps and the ratio of 2/3~:~1/3 in the heights of the steps) are very clear.
}
\end{center}
\end{figure}

The scheme above is realized within gDFTB.\cite{gdftb02,gdftb04-rpp,gdftb04-nl} A
\textit{toy} molecule ($\textrm{C}_{2}\textrm{S}_{2}$) is taken for
the calculation (see Fig.~1). The $\textrm{C}_{2}$ is the CB part,
while $\textrm{S}$ is the linker. The bond length of C-C is
$1.2~\textrm{\AA}$, and the one of S-C is $3.0~\textrm{\AA}$.

In the testing calculation of this chapter, for clarity
within the level structure and the CB characteristics, the
contribution from fragments $\mathbb{C}1$ and $\mathbb{C}2$ is
ignored, and $U_{m}\equiv 1.0~e\textrm{V}$ is taken instead of the
$U_{m}$ calculated by the method in appendix~\ref{app:U} (which are
about $9.0~e\textrm{V}$). Also for simplicity in the calculation, the fictitious
golden leads are used,\cite{Datta02_IEEE} and the minimal basis is
taken.

The transmission function in spectral space is shown in Fig.~5. In
the case without CB correction, we can see that there are seven levels
close to the Fermi energy, while the HOMO and the LUMO+1 are
double-degenerate, respectively. There is no level-split for
charging, though the electronic occupation numbers of the seven
levels above are: 2.00, 1.89, 1.99, 1.99, 0.15, 0.00, 0.00, (it is
clear that LUMO and HOMO-2 are not completely empty/occupied). Then, introducing CB correction,
the LUMO and the HOMO-2 will split into $\epsilon_{m}$ and $\epsilon_{m}+U_{m}$
for the charging effect in the open shell, respectively.

The electronic occupation number of LUMO and the current as a function of bias voltage
$V_{\textrm{bias}}$ are shown in Fig.~6. The charging-induced steps and the correct
ratio of 2/3~:~1/3
in the heights of the steps are very clear, which is consistent
with the results in Fig.~2. The physics under it can be understood with the help
of the charging-induced level-split, which is shown in the following.
1) When $V_{\textrm{bias}}=0$, the LUMO level
is almost empty. 2) Then the positive bias voltage is added, and when the LUMO level
$\epsilon_{\textrm{LUMO}}$ is coming into the Fermi windows
($\mu^{\textrm{R}}_{\textrm{F}}-\mu^{\textrm{L}}_{\textrm{F}}$),
there are two channels
($\epsilon_{\textrm{LUMO},\uparrow}=\epsilon_{\textrm{LUMO},\downarrow}$)
to be opened for current. As leads to the '2/3' in the ratio. 3) After that, the
electronic occupation number of LUMO comes to be 0.66 (see Fig.~6(a)
and Fig.~2(a)), and for the charging effect, the degenerate levels
($\epsilon_{\textrm{LUMO},\uparrow}, \epsilon_{\textrm{LUMO},\downarrow}$) will split
into two ($\epsilon_{\textrm{LUMO},\uparrow}$ and
$\epsilon_{\textrm{LUMO},\downarrow}=\epsilon_{\textrm{LUMO},\uparrow}+U_{\textrm{LUMO}}$).
4) At the time that the level
$\epsilon_{\textrm{LUMO},\downarrow}$($=\epsilon_{\textrm{LUMO},\uparrow}+U_{\textrm{LUMO}}$)
enters the Fermi windows, there will be only one channel to be opened,
which is the reason of the '1/3' in the ratio.

\section{conclusion}

In this paper, we have shown a new way to introduce
CB correction to DFT calculation for the non-equilibrium case.
The main elements of the approach are the following.

1) With the help of the Ansatz in Ref.~\cite{BoSONG_M-CB_06-07},
a self-energy is proposed for the (non-equilibrium) CB
by the model Hamiltonian and EOM approach.
From the comparison with the complete CB results, we can see that
it can include the characteristics of CB correctly.
Further, the more important is that, from the view of DFT-based quantum transport
calculation \textit{in non-equilibrium case},
this self-energy is very convenient for programming more than the
truncation in 
Ref.~\cite{BoSONG_M-CB_06-07} and ME approaches.

2) Based on this self energy, a scheme with DCC is proposed to introduce CB correction
to \textit{non-equilibrium} DFT calculation. 
As is then realized within gDFTB.~\cite{gdftb02,gdftb04-rpp,gdftb04-nl}

By the new code above, the quantum-transport properties of a toy molecule is calculated in CB regime.
In the results of the electronic occupation number and the current as a function of bias voltage,
the CB characteristics (the charging-induced steps and the ratio of 2/3~:~1/3 in the step heights)
appear correctly.

\section{Acknowledgments}
I
thank a lot
Prof.~Gianaurelio~Cuniberti for great supports,
Prof.~Marcus Elstner for helpful discussions, and
Prof.~Thomas Frauenheim and Dr.~Alessandro Pecchia for kindly gDFTB code.

\begin{appendix}
\section{way to calculate $\Sigma^{(\textrm{CB})<}$}
\label{app:Sigma-CB-l} For simplicity, here, only a double-level
($m\equiv 1$ with spin-up and spin-down) case is taken to describe
the quantum dot. Then the index $m$ is ignored in this part, and
from Eq.~(\ref{H-D}), the Hamiltonian of the dot can be re-written
as,
\begin{eqnarray}
H_{\textrm{D}}=\sum_{\sigma}{\epsilon^{0}_{\sigma}d^{\dagger}_{\sigma}d^{\phantom{\dagger}}_{\sigma}}
+\frac{1}{2}\sum_{\sigma}{U n_{\sigma}n_{\bar{\sigma}}}.
\end{eqnarray}
By Eq.~(\ref{GF-u-r}), with the help of the Ansatz in
Ref.~\cite{BoSONG_M-CB_06-07}, we can directly write out
$G^{(U)<}_{\sigma}$ as follows,
\begin{eqnarray}\label{eq:1}
&&G^{(U)<}_{\sigma}=G^{(0)<}_{\sigma}+G^{(0)<}_{\sigma}\Sigma^{a}_{\textrm{H},\sigma}G^{(1)a}_{\sigma}\nonumber+G^{(0)r}_{\sigma}\Sigma^{r}_{\textrm{H},\sigma}G^{(1)<}_{\sigma}\nonumber\\
&&=2\pi\textrm{i}\widetilde{f}(\omega)\delta(\omega-\epsilon^{0}_{\sigma})\left\{1
+\frac{\langle n_{\bar{\sigma}}\rangle U}{\omega-\epsilon^{0}_{\sigma}-U-\textrm{i}\eta}\right\}\nonumber\\
&&\hspace{0.5cm}+\frac{\langle n_{\bar{\sigma}}\rangle U}{\omega-\epsilon^{0}_{\sigma}+\textrm{i}\eta}
\cdot 2\pi\textrm{i}\widetilde{f}(\omega)\delta(\omega-\epsilon^{0}_{\sigma}-U)\nonumber\\
&&=2\pi\textrm{i}\widetilde{f}(\omega)\delta(\omega-\epsilon^{0}_{\sigma})(1-\langle n_{\bar{\sigma}}\rangle)\nonumber\\
&&\hspace{0.5cm}+2\pi\textrm{i}\widetilde{f}(\omega)\delta(\omega-\epsilon^{0}_{\sigma}-U)\langle n_{\bar{\sigma}}\rangle\nonumber\\
&&=\frac{2\textrm{i}\eta \widetilde{f}(\omega)(1-\langle n_{\bar{\sigma}}\rangle)}
{(\omega-\epsilon^{0}_{\sigma})^{2}+\eta^{2}}
+\frac{2\textrm{i}\eta \widetilde{f}(\omega)\langle n_{\bar{\sigma}}\rangle}
{(\omega-\epsilon^{0}_{\sigma}-U)^{2}+\eta^{2}},
\end{eqnarray}
where $\widetilde{f}(\omega)$ is \textit{some kind of pseudo}-Fermi
function in non-equilibrium case, and $\eta\rightarrow 0^{+}$

Assuming that there exists the relation
$\Sigma^{(\textrm{CB})<}_{\sigma}=\textrm{i}\widetilde{f}(\omega)\Gamma^{(\textrm{CB})}(\omega)$,
and $G^{(U)<}_{\sigma}$ can be re-written in the form,
\begin{eqnarray}
G^{(U)<}_{\sigma}=G^{(U)r}_{\sigma}\Sigma^{(\textrm{CB})<}_{\sigma}G^{(U)a}_{\sigma}.
\end{eqnarray}
Then with the help of Eq.~(\ref{GF-u-r}), we will obtain
\begin{eqnarray}\label{eq:2}
G^{(U)<}_{\sigma}(\omega)&=&\textrm{i}\widetilde{f}(\omega)\Gamma^{(\textrm{CB})}(\omega)\left\{
\frac{(1-\langle n_{\bar{\sigma}}\rangle)^{2}}{(\omega-\epsilon^{0}_{\sigma})^{2}+\eta^{2}}\right.\nonumber\\
&&\left.+\frac{\langle n_{\bar{\sigma}}\rangle^{2}}{(\omega-\epsilon^{0}_{\sigma}-U)^{2}+\eta^{2}}
\right\}.
\end{eqnarray}

Comparing Eqs.~(\ref{eq:1}) and~(\ref{eq:2}), we can get
\begin{eqnarray}
\Gamma^{(\textrm{CB})}(\omega)=2\eta D(\omega),
\end{eqnarray}
with
\begin{eqnarray}
D(\omega)=\cfrac{\cfrac{1-\langle n_{\bar{\sigma}}\rangle}
{(\omega-\epsilon^{0}_{\sigma})^{2}+\eta^{2}} +\cfrac{\langle
n_{\bar{\sigma}}\rangle}
{(\omega-\epsilon^{0}_{\sigma}-U)^{2}+\eta^{2}}} {\cfrac{(1-\langle
n_{\bar{\sigma}}\rangle)^{2}}
{(\omega-\epsilon^{0}_{\sigma})^{2}+\eta^{2}} +\cfrac{\langle
n_{\bar{\sigma}}\rangle^{2}}
{(\omega-\epsilon^{0}_{\sigma}-U)^{2}+\eta^{2}}}.
\end{eqnarray}
Therefore, it is clear that
$\Sigma^{(\textrm{CB})<}_{\sigma}=\textrm{i}\widetilde{f}(\omega)\Gamma^{(\textrm{CB})}(\omega)\rightarrow
0$, since $\eta\rightarrow 0^{+}$ and $D(\omega)$ never comes to be
infinite.

%
%

\section{way to calculate Hubbard energy \textit{U} by DFTB/gDFTB}
\label{app:U}

In DFTB/gDFTB,\cite{dftb98,dftb00,gdftb02,gdftb04-rpp,gdftb04-nl,gdftb02-2} the tight-binding Hamiltonian is
written as follows,
\begin{eqnarray}\label{DFTB_H}
H_{\mu,\nu}&=&H^{0}_{\mu,\nu}+H^{\textrm{SCC}}_{\mu,\nu}\\
H^{\textrm{SCC}}_{\mu,\nu}&=&\frac{1}{2}S_{\mu,\nu}\sum^{N}_{\xi}{(\gamma_{\alpha,\xi}+\gamma_{\beta,\xi})
\Delta q_{\xi}}
\end{eqnarray}
with
\begin{eqnarray}
q_{\xi}=\frac{1}{2}\sum^{\textrm{occ}}_{i}n_{i}\sum_{\mu\in\xi}\sum^{N}_{\nu}
{(c^{\ast}_{\mu,i}c^{\phantom{\ast}}_{\nu,i}S_{\mu,\nu}
+c^{\ast}_{\nu,i}c^{\phantom{\ast}}_{\mu,i}S_{\nu,\mu})},
\end{eqnarray}
where $\mu\in\alpha$ and $\nu\in\beta$. $i$ is the index for the
eigenstate of molecule (or molecular fragment), $\mu,\nu$ are the
index of atomic basis, while $\alpha, \beta, \xi$ indicates atoms.
$q_{\alpha}$ is the charge of atom $\alpha$, and $n_{i}$ is the
electron occupation number of eigenstate $i$.
$\gamma_{\alpha,\beta}$ is a function of $U_{\alpha}$, $U_{\beta}$
and $|\mathbf{R}_{\alpha}-\mathbf{R}_{\beta}|$,~\cite{dftb98,dftb00} with $U_{\alpha}$
($U_{\beta}$) is the Hubbard $U$ of atom $\alpha$ (atom $\beta$),
and $|\mathbf{R}_{\alpha}-\mathbf{R}_{\beta}|$ is the distance between
the two atoms. $c_{\mu,i}$ is the project of the eigenvector $i$ on the atomic basis $\mu$,
while $S_{\mu,\nu}$ are the elements of overlap matrix in atomic basis.

With the help of linear combination of atomic orbitals(LCAO) ansatz,~\cite{Jensen99} we can transform the Hamiltonian
matrix~(\ref{DFTB_H}) from atomic basis to molecular basis, and get
the eigenvalue of the eigenstate $i$ as follows,
\begin{eqnarray}
\epsilon_{i}=H_{i,i}=\sum_{\mu,\nu}c^{\ast}_{\mu,i}H_{\mu,\nu}c^{\phantom{\ast}}_{\nu,i}.
\end{eqnarray}

Then, according to the defination of Hubbard $U$,~\cite{LDA+U98,U-def83,U-def05} we
will obtain,
\begin{eqnarray}
U_{i,j}&=&\frac{\partial \epsilon_{i}}{\partial n_{j}}=\frac{\partial H_{i,i}}{\partial n_{j}}=\frac{\partial H^{0}_{i,i}}
{\partial n_{j}}+\frac{\partial H^{\textrm{SCC}}_{i,i}}{\partial n_{j}}\nonumber\\
&=&\frac{\partial H^{0}_{i,i}}{\partial n_{j}}+\sum_{\mu,\nu}\left\{\frac{\partial c^{\ast}_{\mu,i}}
{\partial n_{j}}H^{\textrm{SCC}}_{\mu,\nu}c^{\phantom{\ast}}_{\nu,i}
+c^{\ast}_{\mu,i}H^{\textrm{SCC}}_{\mu,\nu}\frac{\partial c^{\phantom{\ast}}_{\nu,i}}{\partial n_{j}}\right\}\nonumber\\
&&+\sum_{\mu,\nu}c^{\ast}_{\mu,i}\frac{\partial H^{\textrm{SCC}}_{\mu,\nu}}{\partial n_{j}}c^{\phantom{\ast}}_{\nu,i}.
\end{eqnarray}
Ignoring the contribution from $\partial c/\partial n$ and $\partial H^{0}/\partial n$, with
the condition $\partial n_{i}/\partial n_{j}=\delta_{i,j}$, we can get
\begin{eqnarray}\label{U-approx}
&&U_{i,j}\approx\sum_{\mu,\nu}c^{\ast}_{\mu,i}\frac{\partial H^{\textrm{SCC}}_{\mu,\nu}}{\partial n_{j}}c^{\phantom{\ast}}_{\nu,i}\nonumber\\
&&\approx\frac{1}{4}\sum_{\mu,\nu}c^{\ast}_{\mu,i}S_{\mu,\nu}c^{\phantom{\ast}}_{\nu,i}\cdot
\sum^{N}_{\xi}{(\gamma_{\alpha,\xi}+\gamma_{\beta,\xi})}\nonumber\\
&&\cdot\sum_{\mu^{\prime}\in\xi}\sum^{N}_{\nu^{\prime}}
{(c^{\ast}_{\mu^{\prime},j}S_{\mu^{\prime},\nu^{\prime}}c^{\phantom{\ast}}_{\nu^{\prime},j}
+c^{\ast}_{\nu^{\prime},j}S_{\nu^{\prime},\mu^{\prime}}c^{\phantom{\ast}}_{\mu^{\prime},j})},
\end{eqnarray}
with $\mu\in\alpha$, $\nu\in\beta$.

The calculation of Hubbard $U$ by the approximations~(\ref{U-approx})
is performed in four examples: 1)~guanine-cytosine base pair (GC),
2)~adenine-thymine base pair (AT), 3)~benzene
($\textrm{C}_{6}\textrm{H}_{6}$), 4)~double carbon
($\textrm{C}_{2}$) in the case that the bond length is 1.2~$\textrm{\AA}$.
The results are shown in table~I. The error from the approximations is
less than $12\%$, which is acceptable within DFTB/gDFTB.

\begin{threeparttable}\label{tab:U}
\caption{The calculation of Hubbard $U$ by
approximation~(\ref{U-approx})}
\begin{tabular}[t]{lllllllll}
\hline
&~~~~~~& GC &~~~~~& AT &~~~~~& $\textrm{C}_{6}\textrm{H}_{6}$ &~~~~~& $\textrm{C}_{2}$ \\
\hline $U_{\textrm{HOMO}}~(e\textrm{V})$~\tnote{a}
&& 6.170   && 6.235   && 7.157 && 9.040 \\
$\frac{\partial \epsilon_{\textrm{HOMO}}}{\partial n_{\textrm{HOMO}}}~(e\textrm{V})$
&& 5.556   && 5.645   && 6.451 && 9.022\\
\hline $\Delta_{1}~(e\textrm{V})$~\tnote{b}
&& 0.614   && 0.590   && 0.706 && 0.018\\
$\Delta_{2}~(\%)$~\tnote{c}
&& 11.05   && 10.45   && 10.94 && 0.20\\
\hline
\end{tabular}
\begin{tablenotes}
\footnotesize
\item[a] results of HOMO by approximation~(\ref{U-approx}).
\item[b] $\Delta_{1}=U_{\textrm{HOMO}}-\partial \epsilon_{\textrm{HOMO}}/\partial n_{\textrm{HOMO}}$.
\item[c] $\Delta_{2}=\Delta_{1}/(\partial \epsilon_{\textrm{HOMO}}/\partial n_{\textrm{HOMO}})$.
\end{tablenotes}
\end{threeparttable}

\end{appendix}

\newpage

\begin{thebibliography}{37}
\bibitem{note_a}
\bibinfo{note}{Work is in progress to deal with the general case}.
\end{thebibliography}

\end{document}